\documentclass[reprint, amsmath,amssymb,aps, longbibliography]{revtex4-1}%
\usepackage{graphicx, graphics}
\usepackage{wrapfig}
\usepackage[centerlast]{caption}
\usepackage{subcaption}
\usepackage{natbib} 
\usepackage{hyperref}

\usepackage{tabularx}\newcolumntype{Y}{>{\centering\arraybackslash}X}

\usepackage{multirow}

\usepackage{float}

\usepackage{psfrag}
\usepackage{amsmath}
\usepackage{amssymb}
\usepackage{here}
 
\usepackage{fancyhdr}

\begin{document}
	
\title{Value added or misattributed? A multi-institution study on the educational benefit of labs for reinforcing physics content}

\author{N.G. Holmes}
\email[]{ngholmes@cornell.edu}
\affiliation{Laboratory of Atomic and Solid State Physics, Department of Physics, Cornell University, Ithaca, NY}
\author{Jack Olsen}
\affiliation{Department of Physics, University of Washington}
\author{James L. Thomas}
\affiliation{Department of Physics and Astronomy, University of New Mexico}
\author{Carl E. Wieman}
\affiliation{Department of Physics, Stanford University, Stanford, CA}
\affiliation{Graduate School of Education, Stanford University, Stanford, CA}

\date{\today}

\begin{abstract}
Instructional labs are widely seen as a unique, albeit expensive, way to teach scientific content. We measured the effectiveness of introductory lab courses at achieving this educational goal across nine different lab courses at three very different institutions. These institutions and courses encompassed a broad range of student populations and instructional styles. The nine courses studied had two key things in common: the labs aimed to reinforce the content presented in lectures, and the labs were optional. By comparing the performance of students who did and did not take the labs (with careful normalization for selection effects), we found universally and precisely no added value to learning from taking the labs as measured by course exam performance. This work should motivate institutions and departments to reexamine the goals and conduct of their lab courses, given their resource-intensive nature. We show why these results make sense when looking at the comparative mental processes of students involved in research and instructional labs, and offer  alternative goals and instructional approaches that would make lab courses more educationally valuable.
\end{abstract}

\maketitle
\section{Introduction}
Instructional labs have long been considered a mainstay of secondary and post-secondary science courses \cite{Otero2016, Meltzer2015}. In addition to other potential goals, labs are assumed to enhance the learning of course content beyond what is provided by the lectures, recitation sections, and homework \cite{Hofstein1982, Hofstein2004, ALR}. There has been very little evaluation, however, of their actual educational value \cite{Hofstein1982, Hofstein2004, Singer2012, Docktor2014}.  With the ever-greater concerns about college affordability, it is incumbent on university educators to measure what value labs are providing, given their high costs in terms of equipment, space, and instructional staff.

Across the nation, there are calls for science instruction to move beyond traditional content goals to open up the scope of a science education. In K-12 education, for example, the Next Generation Science Standards \cite{Quinn2012} and the revised AP Physics 1 and 2 curriculum \cite{AP2015} have placed significant emphasis on scientific practices and the associated reasoning skills. The American Association of Physics Teachers and the American Physical Society have both published recommendations for undergraduate physics courses to provide students with explicit opportunities to learn transferable skills that will prepare them for a broad array of possible future careers \cite{JTUPP2016, AAPT2014}. These skills include, but are not limited to, developing critical thinking skills, understanding of and proficiency with data analysis, teamwork and collaborative skills, and hands-on experience with physical equipment. Labs offer unique affordances to develop these skills.

Over the last hundred years, the goals of physics education have evolved and cycled. From a focus on explaining everyday physical phenomena, developing the experimental skills of a physicist, or understanding conceptual physics ideas, hands-on activities have frequently been at the core of a physics education \cite{Meltzer2015}. In the late nineteenth and early twentieth century, physics experiments were used as opportunities for students to discover physical phenomena inductively \cite{Otero2016, Meltzer2015}. As curricular requirements became more rigid (to meet college entrance requirements, for example) and enrollment numbers increased, discovery lab activities inevitably evolve to become more structured, leading to well defined experiment protocols \cite{Meltzer2015}, derogatorily referred to as ``cookbook".  These overly-structured labs have been criticized for stifling students' use of cognitive and metacognitive skills \cite{Hofstein2004, Wieman2015CTA, Kung2002, Millar2004}, impeding students' epistemologies regarding the nature of experimentation and scientific measurement \cite{Volkwyn2008, Wilcox2016, Holmes2014}, and being generally ``uninspiring" \cite{Galvez2010}. Nonetheless, step-by-step verification and confirmation labs remain the norm at many institutions, often to satisfy conceptual physics goals by ensuring students will obtain the ``right" answer \cite{ALR, Wieman2015CTA}. Although no one designs labs to be that way, that appears to be a consistent response to external pressures and staffing challenges.

Many institutions and courses, however, are working to move towards more open-ended labs. There are a variety of approaches in doing so. Some instructors have removed all instruction and provide only an experimental goal \cite{Morrison2014}. Others aim to remove some scaffolding \cite{Halstead2016}. Other approaches focus on adding explanation and reflection prompts to increase students' cognition and metacognition \cite{Kung2002, Kung2007}. Some courses, such as workshop or studio physics \cite{Laws1991, Laws2015, Cummings1999, Beichner2007, Beichner2000}, have removed the divisions between lecture, lab, and recitation and incorporated all three aspects in to a single learning environment (see also Investigative Science Learning Environments, e.g. \cite{Etkina2007}). In these cases, hands-on activities are used fluidly with other learning activities as students discover conceptual ideas. In many of these cases, there was still a significant aim to support students' understanding of physics content. 

Recent work probing students' attitudes towards and beliefs about experimental physics have raised concerns about this goal \cite{Wilcox2016, Wilcox2017}. In these two studies, almost 5000 students in over 100 courses at 67 different institutions were surveyed at the start and end of a physics lab course about several aspects of experimental physics. The researchers compared the responses of students based on the structures of the lab courses in which they were enrolled. They found that students' perceptions of experimental physics became less expert-like in courses that were more guided compared with ones with open-ended activities \cite{Wilcox2016} and in ones whose primary goals included focus on developing or reinforcing physics concepts compared with ones that focused on skills \cite{Wilcox2017}. 

In line with these studies, our own previous work found that two moderately structured lab courses, designed to reinforce physics content knowledge with significant focus on student reflection and reasoning, did not provide measurable added-value to students' learning \cite{Wieman2015AJP}. That single-institution, two-course study was limited in its generalizability, particularly because of the highly selective nature of the institution. 

In response to these issues, and the general lack of controlled research studies evaluating pedagogies and curricular approaches to physics lab courses, we have extended that work to nine instances of six different university-level courses taught at three institutions encompassing a very diverse set of students. In all cases, the labs had the same goal: to reinforce the physics content presented in lecture. 

We had three research questions in this study. Our first asked: what is the impact of taking a lab course on student learning as measured by student performance on exams in the associated lecture course? One may argue, however, that typical physics exams primarily ask quantitative problem solving questions. The effects of the labs, particularly the notion of seeing and experiencing the physics content, may be unlikely to provide benefit on such questions. Instead, perhaps the lab courses benefit conceptual learning. Our second research question, therefore, asked whether lab courses selectively impacted the learning of concepts. Finally, our third research question asked whether there might be short-term learning benefits that get washed out on the final exams.

In what follows, we demonstrate with a high degree of precision and complete consistency across courses and institutions that labs are not adding measurable value to students' course performance. We will conclude with a number of suggestions from physics education research for how to restructure lab courses to emphasize other skills and scientific practices.

\section{Methods}

To answer all three research questions, we first needed to isolate the impact of the lab from the remainder of the course. At all three institutions, the lab courses were associated with an introductory physics lecture course, though enrolling in the lab courses was optional. In each case, a subset of the lecture students also enrolled in the associated lab courses. This allowed us to isolate the impact of the lab by comparing the performance of lab-enrolled and non-lab-enrolled students. 

\subsection{Sample populations}
The first institution was a public research university in southwestern United States. Approximately 55\% of the enrolled undergraduate students at the institution are female and the most prevalent races/ethnicities are Hispanic (47\%) and White (34\%). Upon entering the institution, students' average high school GPA is 3.39 \cite{UNM}. The courses studied at this institution were both part of a calculus-based introductory physics sequence with three hours of lecture each week and optional recitation sessions. The lab sessions were three hours each week. The courses primarily served engineering majors at the sophomore or junior level. 

The second institution was a public research university in northwestern United States. Approximately 52\% of the enrolled undergraduate students are female and the most prevalent races/ethnicities are Caucasian (41\%) and Asian (25\%). Upon entering the institution, students' average high school GPA is between 3.64 and 3.93 \cite{UW1}. Both the courses studied at this institution were part of an algebra-based introductory physics sequence with four hours of lecture each week. The lab sessions were two hours each week. The courses primarily served health science majors at the senior or junior level. 

The third institution, which is the same as that used in reference \cite{Wieman2015AJP}, was an elite, private research university in southwestern United States. Approximately 50\% of the undergraduate students are female and the most prevalent races/ethnicities are White (40\%) and Asian (20\%). Upon entering the institution, students' average high school GPA is 4.00 or above \cite{SU1, SU2}. The courses studied at this institution were both part of a calculus-based introductory physics sequence with three hours of lecture and two hours of recitation each week. The lab sessions were two hours each week. The courses primarily served engineering majors at the freshman and sophomore level. The two courses used in \cite{Wieman2015AJP} and one used in \cite{Wieman2015PERC} make up this data set. 

Additional information about the institutions, courses, and student populations can be found in the Appendix.

\subsection{Data and analysis}
We used student scores on the final course exams as a performance measure of learning the course material. All topics on the final exams were covered in the lecture and/or recitation sections. The lab courses, however, only covered a subset of the course topics. We categorized each exam question as to whether it was lab related (meaning there was an associated lab activity that covered all the content needed to fully answer the question) or non-lab related (meaning there was no associated lab activity that covered all the content needed to fully answer the question). To answer the second research question, we also categorized each exam question as to whether it involved primarily conceptual or calculational reasoning (the necessary exam information was only available for Schools 1 and 2). Conceptual questions were defined as ones that did not require any calculation and could be answered simply by reasoning qualitatively about the physical systems. The Appendix include examples of these categorizations.

If the students in the two groups (lab enrolled or non-lab enrolled) were equivalent, we could then simply compare the performance of lab-enrolled and non-lab-enrolled students on the lab-related questions to determine the added value of the labs (using either the full exams or the conceptual items only). Students who opted to enroll in the optional lab courses, however, were distinctly different from those who did not. Lab-enrolled students typically outperformed the non-lab-enrolled students on exams and, when the data were available, scored higher on pre-course measures of physics knowledge. Some majors (such as physics at School 1 and engineering physics at School 3) required the lab courses for graduation. 

To account for these selection effects, we calculated a difference score for each student, defined as the difference between the students' fractional score on the lab-related (ScoreL) and non-lab-related (ScoreNL) questions, as in \cite{Wieman2015AJP}. This difference provides a measure of the student's relative performance on the two types of questions (Equation \ref{eqn:DS}). For example, a student who scores 1 (i.e. 100\%) on the lab-related items and 0 on the non-lab-related items would get a difference score of 1. A student who scores 0.5 (i.e. 50\%) on the lab-related items and 0.75 (i.e. 75\%) on the non-lab-related items would get a difference score of -0.25. 

\begin{equation}
difference \, score = Score_L - Score_{NL}
\label{eqn:DS}
\end{equation}

By comparing the average of these difference scores between the lab-enrolled and non-lab-enrolled students, we obtained a measure of the added value of the labs. This compensates for student selection effects and variations in the difficulty of the test items within and between exams and courses. We define the mean lab benefit in Equation \ref{eqn:MLB}, (where NLab is the number of lab-enrolled students and NNon-Lab is the number of the non-lab-enrolled students), as in \cite{Wieman2015AJP}. The null hypothesis is that the mean lab benefit is zero.

\begin{multline}
mean \: lab \: benefit \\ = \frac{1}{N_{Lab}} \sum_{i=1}^{N_{Lab}} \left(difference \: score\right)_i - \\ \frac{1}{N_{Non-Lab}} \sum_{j=1}^{N_{Non-Lab}} \left(difference \: score\right)_j 
\label{eqn:MLB}
\end{multline}

\section{Results}

\begin{figure*}
\begin{center}
\includegraphics[width=0.9\textwidth]{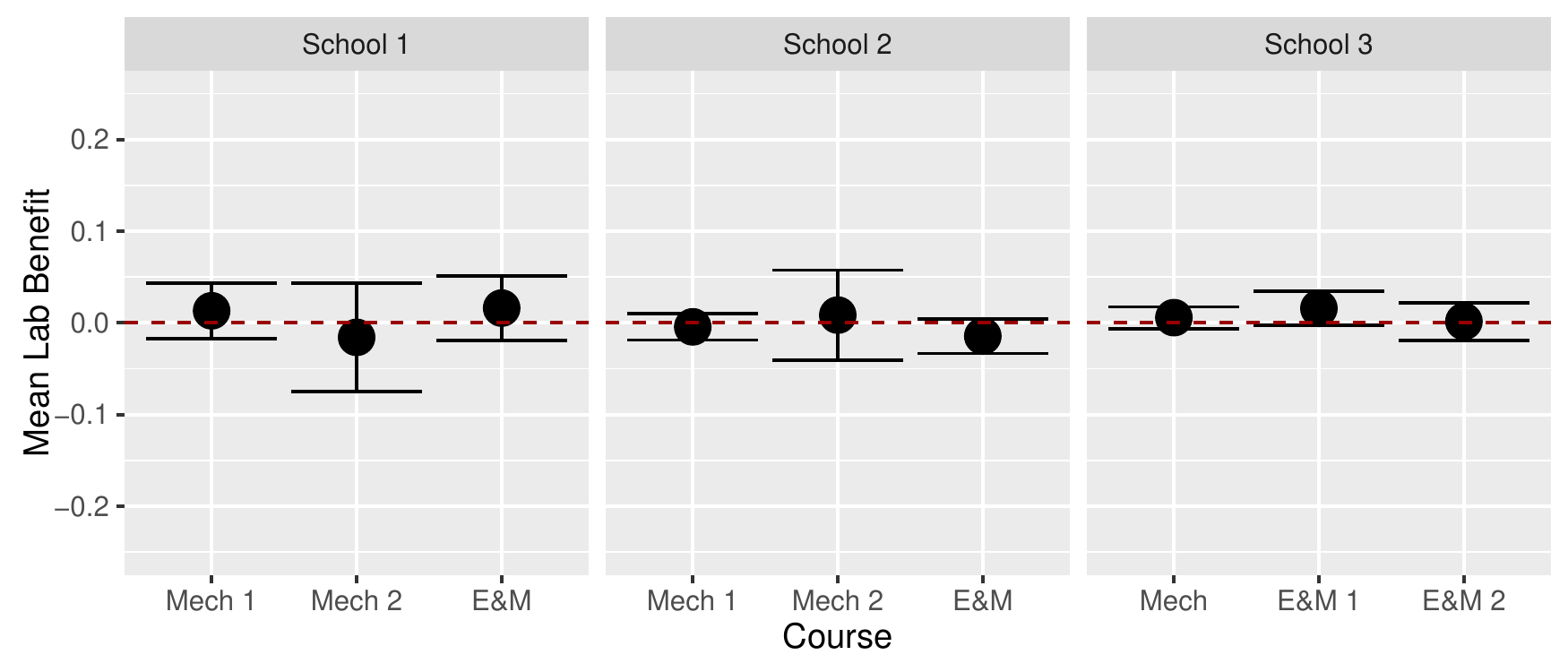}
\caption{The figure shows the mean lab benefit (Equation \ref{eqn:MLB}) on the final exams in nine courses at three institutions. Error bars represent the standard error (= standard uncertainty in the mean). The dashed line indicates the hypothesis being tested (whether the mean lab benefit differs significantly from 0).}
\label{fig:Finals}
\end{center}
\end{figure*}

To evaluate the first research question, we compared the mean lab benefit for all final exams in the courses at the participating institutions. From Fig. \ref{fig:Finals}, we see that the mean lab benefit is consistent with zero across all nine courses. Combining all the data, we find that the average mean lab benefit across courses was found to be: $0.0024 \pm 0.0067$ (or $0.24 \pm 0.67 \%$). An ANOVA comparing the differences between lab-enrolled and non-lab-enrolled students on their difference scores was non-significant: $F(1,2615) = 2.32, p = .128$ (for statistical analysis see Table \ref{tab:FullStats}).

\begin{figure}
\begin{center}
\includegraphics[width=0.5\textwidth]{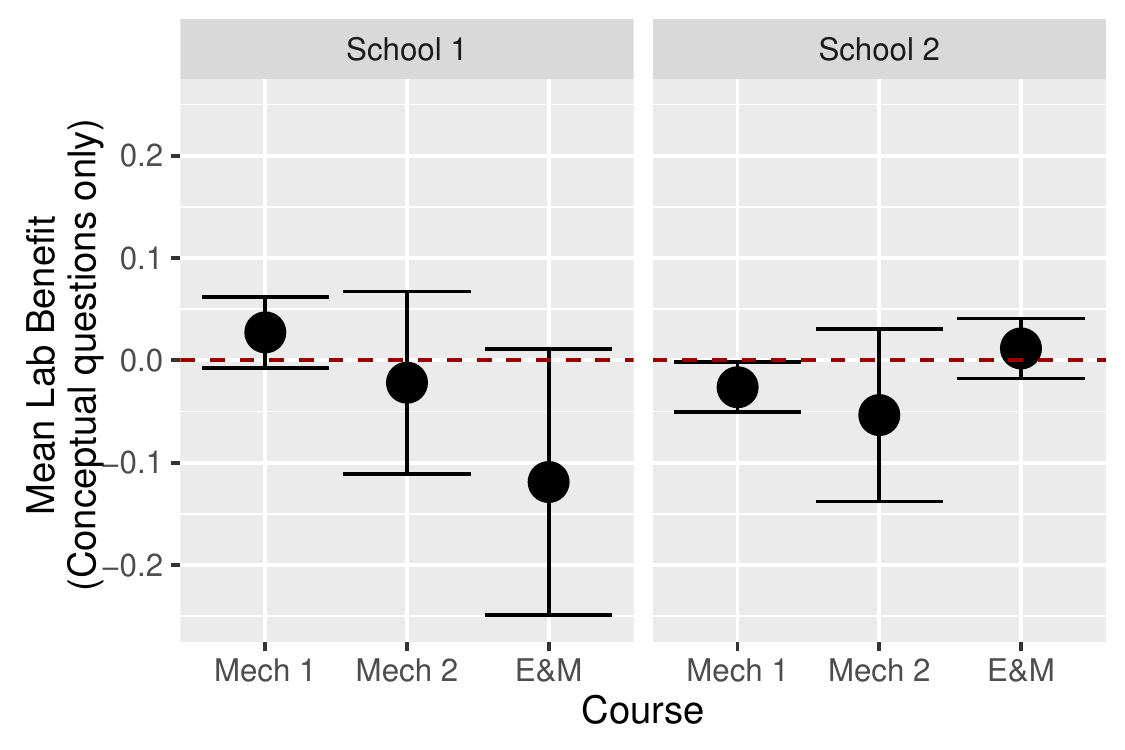}
\caption{The figure shows the mean conceptual lab benefit (Equation \ref{eqn:MLB}) for six courses at two institutions. Error bars represent the standard error (= standard uncertainty in the mean). The dashed line indicates the hypothesis being tested (whether the mean lab benefit differs significantly from 0).
}
\label{fig:FinalsC}
\end{center}
\end{figure}

The second research question asked whether the impact of the labs occurred only for conceptual reasons. We, therefore, calculated the mean lab benefit as before, but including only the conceptual exam items. Again, we see no significant differences from zero for any course (Fig. \ref{fig:FinalsC}). The average mean conceptual lab benefit across courses (all data combined) was found to be: $-0.0062 \pm 0.0158$ (or $-0.62 \pm 1.58 \%$). An ANOVA comparing the differences between lab-enrolled and non-lab-enrolled students on the difference scores for conceptual items was non-significant: $F(1,1114) = 1.60, p = .206$ (Table \ref{tab:FullCStats}).

\begin{figure*}
\begin{center}
\begin{subfigure}{0.8\textwidth}
\includegraphics[width=\textwidth]{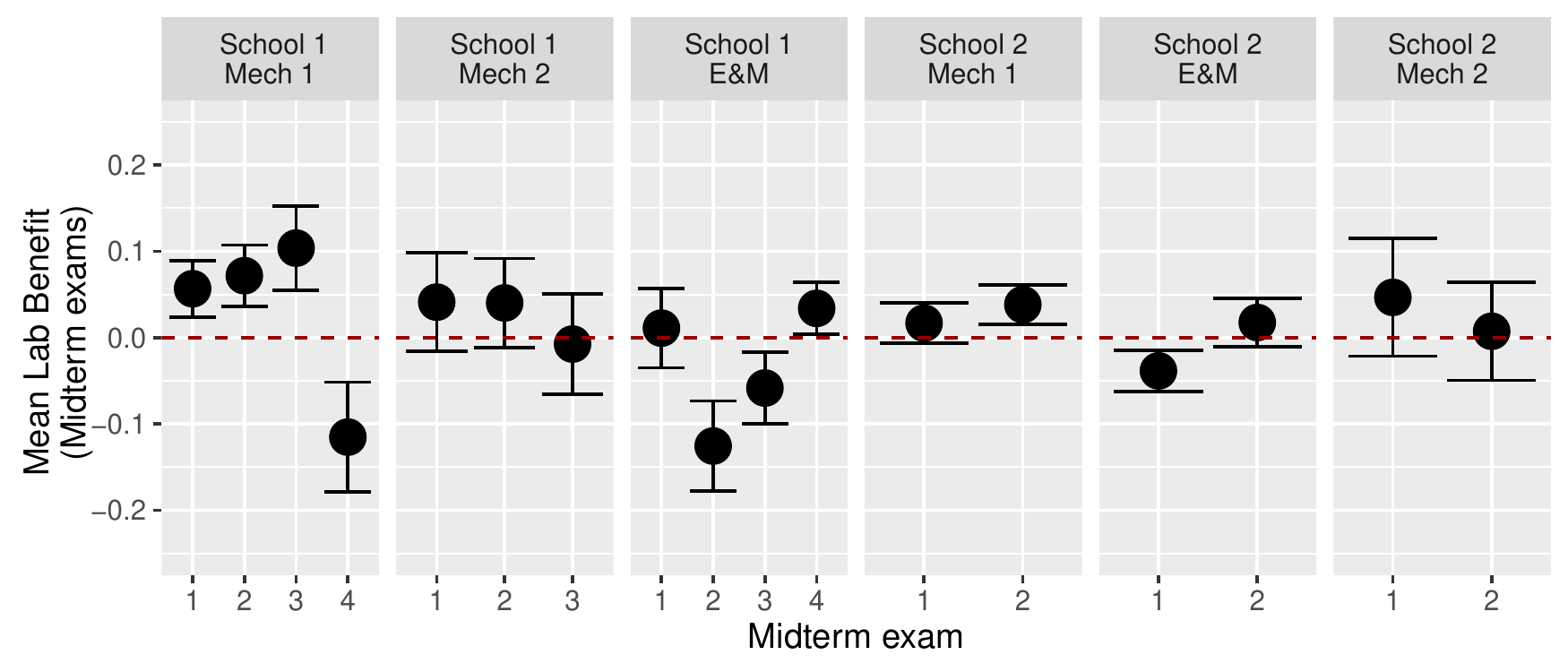}
\caption{Difference scores on the full test across midterm exams. }
\label{fig:Midterms}
\end{subfigure}

\begin{subfigure}{0.7\textwidth}
\includegraphics[width=\textwidth]{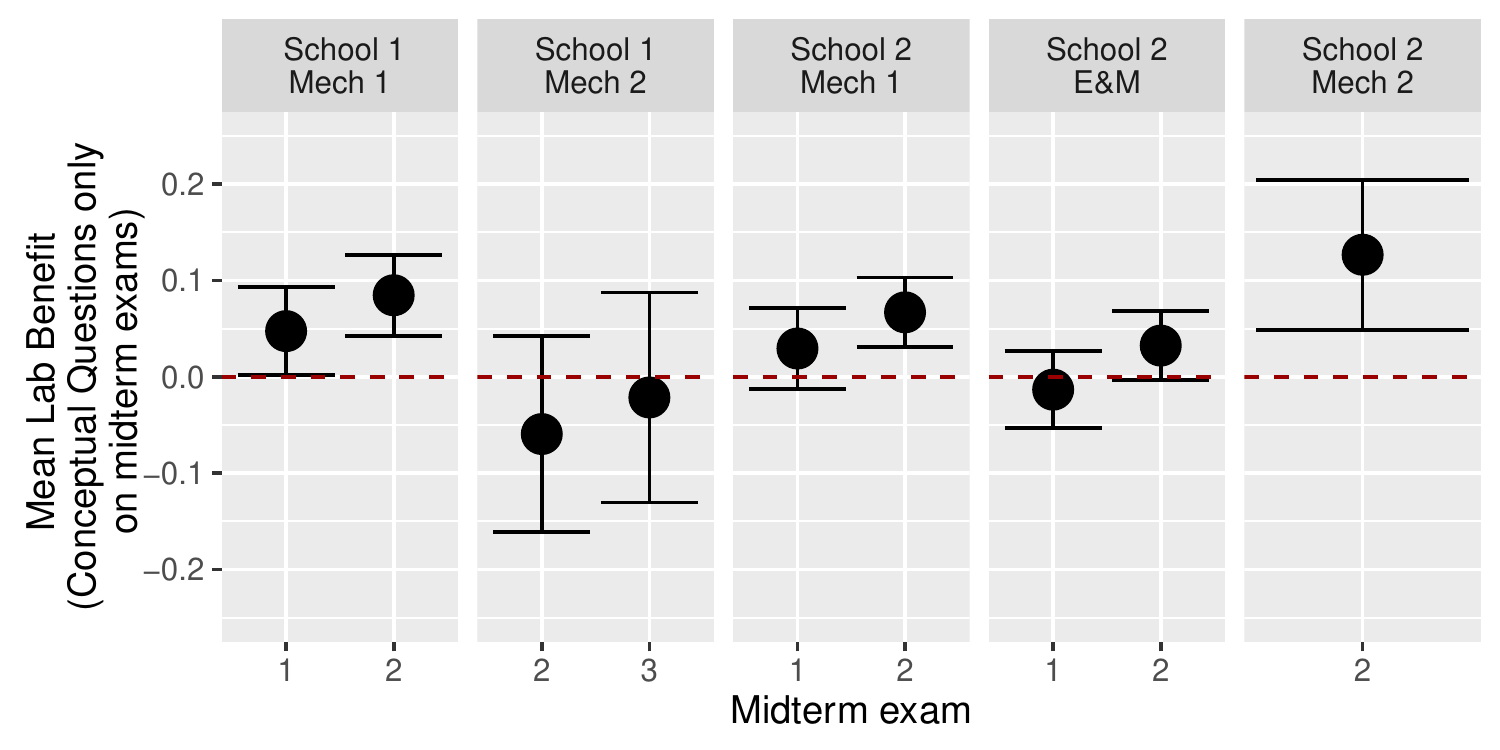}
\caption{Difference scores on the conceptual-items only across midterm exams. }
\label{fig:MidtermsC}
\end{subfigure}
\caption{The figure shows the mean lab benefit (Equation \ref{eqn:MLB}) across midterms for six courses at two institutions for a) the full exams and b) the conceptual items only. Not all midterm exams had sufficient conceptual questions to be included in the analysis. Error bars represent the standard error (or standard uncertainty in the mean). The dashed line indicates the hypothesis being tested (whether the mean lab benefit differs significantly from 0).}
\end{center}
\end{figure*}

Finally, we evaluated whether short-term benefits may have washed out by the final exam. We used the characterizations of the questions (lab-related or non-lab-related and conceptual or calculational) for all midterm exams in the six courses at Schools 1 and 2 for which data was available (see Table \ref{tab:MidStats} and Fig. \ref{fig:Midterms} and \ref{fig:MidtermsC}). Repeated-measures ANOVA analyses comparing the difference between lab-enrolled and non-lab-enrolled students on their difference scores was again non-significant: for the full exams, $F(1,2939) = 0.65, p = .420$; and for the conceptual questions only, $F(1,2050) = 0.10, p = .757$. Combining all the data, the mean lab benefit for the full exams was $0.0082 \pm 0.0111$ (or $0.82 \pm 1.11 \%$) and for the conceptual questions only was $0.033 \pm 0.022$ (or $3.3 \pm 2.2 \%$, note that this measure is less reliable than repeated-measures ANOVA analysis for the midterm exams due to the re-occurrence of the same students multiple times in the data set).

\section{Discussion}

Across many different exams in many different courses at three different institutions we found no measurable learning benefits of the lab courses, often with strikingly high precision. Although this result may seem surprising, given the prevalence of such instructional labs with these goals, there is some theoretical rationale. First, goals to reinforce content often come hand-in-hand with increased structure, as it becomes important for students to observe a particular `correct' result \cite{Meltzer2015}. When one examines the cognitive activities in which students are engaged while completing such lab course activities \cite{Wieman2015CTA, Holmes2016}, they are dominated by following instructions to collect specified data using unfamiliar equipment, and following specified procedures to analyze the data and write up reports in a specified format. Although the relevant physics concepts were central to the thinking of the instructor that designed and built the experiments, those concepts get little, if any, attention from the student carrying out the assigned activities using that apparatus. 

Although there are some theoretical arguments for how labs could contribute to learning, such as the cognitive benefits of multiple representations \cite{Moreno1999, Ainsworth2002}, physical manipulatives \cite{Martin2005, Tsang2015, Schwartz1999}, or embodied cognition \cite{Anderson2003, Wilson2002}, there is little in the design of these lab activities to realize such potential benefits. 

There are several limitations of this work. We compared the isolated effect of labs when lecture and recitation sections were also taking place, albeit distinctly. In addition, many of these courses included extensive opportunities for students to make predictions and reflect on the physics concepts at hand. In most cases, the lab activities occurred after instruction on the associated content. Any predictions, therefore, would be made from previously learned concepts or equations. We cannot claim that labs such as these cannot or do not teach content knowledge. What we have shown is that taking the labs is demonstrating no measurable added value beyond what is achieved by other aspects of these courses. 

The mean lab benefit measure is itself limited in that it only provides an average glimpse across the class and does not reveal whether there are subgroups of students for whom the lab may have an effect. If there are systematic effects for subpopulations, the average null result would suggest that the lab benefits some students but negatively impacts others. Arguably, this may be a more problematic scenario than concluding that there is no effect. 

Another potential criticism of this method is that the we have assumed that the lab activities will preferentially benefit learning the lab-related content. Some may argue that the learning that takes place in lab may transfer to other topics in the course. Research on transfer would suggest that this is rather optimistic \cite{Bransford1999, Nokes-Malach2013}. It is also inconsistent with the low correlation between answers to different questions on the same exam that we observed previously \cite{Wieman2015AJP}. Another possible claim is that the lab increases student motivation, which then is reflected in improved overall course performance. Studies of both concepts-based and structured lab courses, however, have shown that student attitudes and motivation towards experimental physics decrease by the end of such courses \cite{Wilcox2016, Wilcox2017}. Regardless of whether the learning in the lab leads to improved learning in other aspects of the course, what we have shown is that these labs are not achieving their stated goals. The goals of these courses were to help develop or reinforce student understanding of the relevant physics material. Our study has shown that this goal is not being achieved to a measurable extent. 

In addition, all of this work dealt with introductory physics labs, and therefore it is unclear whether upper-division lab courses would produce the same results. The work evaluating student beliefs about experimental physics did examine introductory and beyond-first-year labs separately and found the same effects (guided and concepts-focused labs both resulted in less expert-like beliefs). Our study, therefore, emphasizes the need to rigorously evaluate how well those labs are achieving their educational goals.

We also note that ``studio" physics courses (where labs, lectures, and recitation are co-mingled in the same space and instructors move fluidly between the three forms of instruction) have demonstrated significant conceptual learning gains compared to traditional instruction \cite{Docktor2014, Laws2015, Cummings1999, Wilson1994}. It is impossible to separate the contributions of the labs in this integrated context, but they may provide greater benefits when used in this way. 

Nonetheless, distinct lectures, recitations, and labs, like the courses examined here, remain the norm at many large institutions due to practical and logistical constraints. The consistent pattern we see of zero added value for learning the course content demands that instructors and departments critically evaluate the goals and outcomes of their instructional labs, particularly given their high cost.  

In line with many national calls \cite{Quinn2012, AP2015, JTUPP2016, AAPT2014}, our results should strongly encourage institutions to consider alternative goals and pedagogies for labs, especially those for which labs are demonstrably and uniquely effective. There is much research demonstrating the effectiveness of lab-based pedagogies at developing skills such as evaluating data and models, dealing with uncertainty and variability in data, and designing experiments \cite{Holmes2015, Etkina2010, Zwickl2014, Lewandowski2015, Pillay2008, Buffler2008}.

One need not completely discard introductory lab activities to begin to make improvement. Research has shown that even small elements of open-endedness in activities can improve student attitudes towards experimental physics \cite{Wilcox2016}. Providing students with time, opportunity, and incentive to revise, troubleshoot, or explore by, for example, spreading a single lab experiment across multiple weeks may enable the desired skills focus \cite{Holmes2016, Holmes2015}. Shifting the emphasis of the lab activities towards the quality of students' process rather than the product they obtain would be key to facilitating that development \cite{Holmes2016}. It is unclear, however, whether entirely open-ended project-based courses would achieve the desired goals, especially when scaled for typical introductory course enrollments. Indeed, history has suggested that this is not a sustainable framework \cite{Meltzer2015}. 

\section{Conclusion}

Our results have demonstrated a broken link between intended learning goals and measures of student outcomes. Nine different lab courses, designed to reinforce student understanding of physics content from other areas of the course, have been shown to provide no measurable added value to course performance. This was true across calculus-based and algebra-based courses at three very diverse institutions. We hope these results will encourage instructors and departments to critically evaluate whether their lab courses are achieving their full potential. 

This work motivates many new research areas and questions, the most important of which is how to provide educationally effective experimentation experiences to introductory physics students.

\acknowledgements
We would like to thank the instructors across all three institutions for providing the data, S. Salehi for consultation of data analysis methods, and P. Heron for her support.

\textbf{Author contributions:} N.G. Holmes analyzed the data and wrote the manuscript. N.G. Holmes and J. Olsen categorized each exam question. J. Olsen and J. Thomas collected and anonymized all data. C.E. Wieman supervised the work did extensive editing of the manuscript. All authors edited the manuscript.

\bibliography{LNLPaper}

\appendix*

\section{Data sources and analysis methods}

In the following sections, we provide more extensive descriptions of our data sources and analysis methods.

\subsection{Data sources and student populations}

\begin{table*}
\caption{The participants at each institution and course are broken down by the number of students enrolled in the lab (\# Lab SS) and not enrolled in the lab (\# Non-Lab SS). The associated exams are described by the number of lab- and non-lab-related items (Qs), conceptual items (CQs), and quantitative items (QQs). Note that some items (such as factual recall questions) were not categorized as conceptual nor quantitative.}\label{tab:Peeps}
\begin{tabularx}{\textwidth}{ccc|YY|YYYYYYY}
\hline\hline
School & Course & Exam & \# Lab Ss & \# Non-Lab Ss & \# Items & \# Lab Qs & \# Non-Lab Qs & \# Lab CQs & \# Non-lab CQs & \# Lab QQs & \# Non-Lab QQs\\
\hline
\multirow{14}{*}{School 1} & \multirow{5}{*}{Mech 1} & 1 & 71 & 129 & 15 & 6 &9 & 6 &3 & 0 & 6 \\
&& 2 & 72 & 120& 12& 5 & 7&4&4&1&3\\
&& 3 &60&86& 12& 4 &8 &3& 1&1 &6\\
&&4 &57&72&10&1&9&1&6&0&3\\
&&Final &57 & 67&30&12&18& 8& 10 &4&8\\
\cline{3-12}

& \multirow{4}{*}{E\&M} & 1&25 &63 &15& 4& 11& 1& 3& 3&8\\
&& 2 & 22 &56& 15& 4 &11 &2 &2& 2& 9\\
&& 3 &20 & 53 & 16& 7 &9 &2&3&5&6\\
&& Final & 18 &53& 22 &9 & 13 &4 &3 &5 &10\\
\cline{3-12}
& \multirow{5}{*}{Mech 2}& 1 &34 & 80 & 17 & 3 &14 & 1 &8 & 1 & 6\\
&& 2 & 37 & 85 & 17 & 3 & 14 &0 &1 & 3 &13\\
&& 3 & 34 &76 & 18 & 6 & 12 & 2 & 1 & 4 &10\\
&& 4& 31 & 73 & 25 &8 &17 & 3 &2 & 4 &15\\
&& Final & 37 & 80 & 27 & 7 & 20 & 1 &1& 6 & 17\\

\hline

\multirow{12}{*}{School 2} & \multirow{4}{*}{Mech 1} & 1 & 231 & 173 & 15 & 7 & 8 & 3 & 2 & 4 & 6\\
&& 2 & 221 & 168 & 15 & 6 & 9 & 3 &3 &3 & 6\\
&& Final A & 114 & 114 & 40 & 15 & 25 & 7 & 6 & 8 & 19\\
&& Final B & 112 & 112 & 40 & 14 & 26 & 8 &8 & 6& 18 \\
\cline{3-12}
& \multirow{5}{*}{E\&M} & 1& 183& 140& 15& 6& 9& 2& 4& 4& 5\\
& & 2-A& 105& 78& 15& 4& 11& 4& 3& 0& 8\\
& & 2-B& 76& 59& 15& 3& 12& 3& 4& 0& 8\\
& & Final-A& 105& 78& 40& 9& 31& 4& 11& 5& 20\\
& & Final-B& 80& 65& 40& 10& 30& 5& 11& 5 & 19\\
\cline{3-12}
& \multirow{3}{*}{Mech 2} & 1& 49& 32& 15& 9& 6& 4& 1& 5& 5\\
& & 2& 48& 33& 15& 7& 8& 4& 3& 3& 5\\
& & Final & 49& 33& 20& 8& 12& 2& 8& 6& 4\\
\hline
\multirow{3}{*}{School 3} & Mech &  Final &  211& 360& 0& 6& 14\\
\cline{3-8}
& E\&M 1 & Final & 129 & 361& 20& 11& 9\\
\cline{3-8}
& E\&M 2 & Final & 126& 317& 20& 7& 13\\
\hline\hline
\end{tabularx}
\end{table*}

Data sources included in the study came from three different institutions and three courses at each institution. All courses were introductory-level physics courses covering mechanics (Mech) or electricity and magnetism (E\&M). Some courses were different instances of the same course (that is, in a different semester and/or with a different instructor). There were nine different, unique courses in the analysis in total. At Schools 1 and 2, we used only the multiple-choice exam questions to reduce variability from multiple graders across multiple exams. Table \ref{tab:Peeps} characterizes the number of students at each institution and the details of the exams used. The demographics and course details described in the following sections differs between institutions based on the availability of information.

\subsubsection{School 1}
The first institution was a public research university in the southwestern United States. Approximately 55\% of the enrolled undergraduate students at this institution are female, less than 2\% are international students, 47\% are Hispanic, 34\% are White, 6\% are American Indian, 4\% are Asian, 2\% are African American, less than 1\% are Native Hawaii'an, and the remaining are two or more races or unknown race/ethnicity. Upon entering the institution, students' average high school GPA is 3.39 and their average SAT Math score is 540.93 \cite{UNM}.

Two instances of the same introductory mechanics course with different instructors and one introductory electricity and magnetism course were included in the analysis. Both courses were part of the same introductory calculus-based physics sequence. The institution also offered an algebra-based sequence for pre-medical students and an algebra-based sequence for non-science majors. Around 100 students enroll in these courses each semester and a third to half of the students also take the lab course. Students enrolled in the mechanics lecture course were primarily engineering or computer science majors (78\%) at the sophomore level (43\%; 37\% at the junior-level, 14\% at the freshman-level). Nearly 70\% of the students were male. Of the students also enrolled in the lab course, there was a slightly higher fraction of male students (77\%) and fewer engineering or computer science majors (69\%) compared to the lecture alone. There was also a more even distribution of freshmen (31\%), sophomores (36\%), and juniors (33\%). The students in the electricity and magnetism course follow a similar demographic distribution.

All courses in this sequence involved three hours of lecture each week with optional recitation sessions. The lab courses involved weekly three-hour lab sessions. Each lab was facilitated by a graduate teaching assistant (TA). Students purchased a published lab notebook that includes prompts and questions to guide them (in groups of two or three) through each experiment. The guide provided space for predictions, data collection, and reflection. There were no pre-lab questions, though most labs included a post-lab reflection or synthesis. The students were graded based on their performance on the in- and post-lab questions. 

The mechanics lab course primarily used carts, track, motion sensors, and force sensors (carts-on-track equipment) to explore the concepts of motion, force, energy, momentum, torque, periodic motion, and waves in pace with the lecture. The electricity and magnetism lab course began with heat and pressure concepts and then transitioned into electric force, field, potentials, and circuits, and magnetism. In both cases, the labs focused primarily on reinforcing or introducing topics associated with the lecture. There were some activities targeting uncertainty, graphical, and data analysis, mathematical procedures (such as vector analysis), and equipment fluency. 

\subsubsection{School 2}

The second institution was a public research university in the northwestern United States. Approximately 52\% of the enrolled undergraduate students at this institution are female, 14\% are international students, 41\% are Caucasian, 25\% are Asian, 7\% are Latino, 7\% are two or more races, 3\% are African American, 0.5\% are Hawaii'an/Pacific Islander, 0.4\% are American Indian/Alaskan Native, and the remaining are other or not indicated \cite{UW2}. Upon entering the institution, students' average high school GPA is between 3.64 and 3.93 and their average SAT Math score is between 570 and 690 \cite{UW1}. 

Two instances of the same mechanics course and an electricity and magnetism course were included in the analysis, all taught by the same instructor in different terms. Both courses were part of the same introductory algebra-based physics sequence. The institution also offered a calculus-based sequence for physical science and engineering students, as well as courses for non-science majors. Around 400 students were enrolled in each instance of these courses and around 60\% of the students also enrolled in the lab course. The courses were comprised mostly of students pursuing Biology majors (55\%), with a significant proportion of students pursuing health-related (12\%) or other science majors (11\%). Most students were seniors (54\%) or juniors (33\%). Most students hailed from the US (over 90\%) and nearly 80\% of these were in-state. Around 60\% of the students were female. The students were predominantly Caucasian (40\%) or Asian (30\%). The average age of the students was 20.43 years, with the majority aged either 20 or 21.

The lecture courses involved four hours of lecture each week led by the faculty instructor. Each lab course consisted of eight two-hour-long in-person lab sessions led by a TA. Each lab session included online pre-lab and post-lab activities. Students purchased a published lab notebook with prompts and questions that guide them (in groups of two or three) through each experiment. The lab manual provided space for predictions, data collection, and reflection. Interspersed throughout each lab were ``TA Check-off" moments, where the students were required to discuss their progress with the TA and receive a mark in their lab manual. Students were required to pass all ``TA Check-off" moments to complete the lab. The students were graded based on their performance on the pre- and post-lab questions and on their participation in the in-person session (not just for their physical presence).

The mechanics labs employed the carts-on-track equipment liberally to explore motion, forces, energy, momentum, and rotational motion in pace with the lecture. There were 41 explicit goals, 61\% of which focused on conceptual fluency, 27\% of which focused on data acquisition and processing, and 12\% of which focused on equipment fluency. The electricity and magnetism lab course began with heat concepts, transitioned into electric charge, field, potential, and circuits concepts, and finished with magnetism. There were 28 explicit goals, 61\% of which focused on conceptual fluency, 29\% of which focused on data collection and interpretation, and 11\% of which focused on equipment fluency. There was a third course in the sequence (waves and optics) that was not examined in this study.

\subsubsection{School 3} 

The third institution was an elite, private institution in the southwestern United States. At the institution, just under 10\% of students are international, and almost 40\% of students come from in-state. Approximately 50\% of students identify as female, 40\% of students identify as white/Caucasian, 20\% as Asian, 8\% African-American, 6\% Mexican/Chicano, 6\% other Hispanic, 2\% Native American, 1\% Native Hawaiian/Pacific Islander, and the remaining other or unknown \cite{SU2}. More than 75\% of the admitted (first-year) class have a high school GPA of 4.0 or above and an SAT Math score over 700 \cite{SU1}. The data from this institution were used previously in \cite{Wieman2015AJP, Wieman2015PERC}.

One mechanics course and two instances of the same electricity and magnetism course were included in the analysis. The same instructor taught both instances of the electricity and magnetism course. Both courses were part of the same introductory calculus-based physics sequence. Around 500 students were enrolled in each course and about 30\% of the students also enrolled in the lab course. Students in these courses were primarily engineering majors, with a minority of students intending to pursue physics, medicine, or other scientific disciplines. The institution also offered an honors sequence for physics majors and an algebra-based sequence for pre-medicine students. Approximately 50\% of the students were freshman (first-year) and 30\% were sophomores (second-year). Just under 40\% of the students were female.

The courses involved three one-hour lectures led by a faculty instructor and a one-hour TA-led recitation session each week. The lab courses involved a two-hour TA-led session with a small pre-lab activity each week. Lab guides were posted on the course management system and students were expected to print and bring them to each lab session. The lab guides provided space for predictions, data collection, and reflection. Students were expected to stay in the lab until they had answered all questions in the lab guide. Completed pre-lab and in-lab guides were submitted at the end of each lab session and were evaluated by the TA. The lab course, however, was pass or fail based on participation (not just for their physical presence).

The mechanics labs primarily employed carts-on-track equipment to explore motion, forces, energy, and momentum in pace with the lecture portion of the course. Each lab listed one or two explicit goals almost exclusively focused on conceptual fluency (with only one goal focused on equipment fluency). The electricity and magnetism lab course included lab activities about electric forces, fields, potentials, and circuits, and magnetism. In the first instance of the lab course, each lab listed one or two explicit goals primarily about conceptual fluency, with several explicit instances focused on data analysis (such as how to choose appropriate graphs to represent data) and equipment fluency (including an entire lab session about instrumentation). In the second instance of the lab course, lab activities were redesigned to also include learning goals associated with critical thinking, data analysis, and experimental design as in \cite{Holmes2015}. There was a third course in the sequence (light and heat) that was not examined in this study.

\subsection{Categorization methods}
Two raters categorized each exam question according to whether it was related to a lab activity and whether it was primarily conceptual. The two raters then compared their categorizations and any disagreements were discussed and resolved. On the first round, there was greater than 75\% agreement on the categorizations. One of the raters was a coordinator of the labs at one institution and, thus, also provided on-the-ground expertise regarding whether items were reasonably related to the lab activity. A third rater, who was an instructor at the second institution, evaluated a subset of the completed categorizations for that institution to provide similar expertise check. All disagreements were discussed and resolved on consensus. 

\begin{figure}
\begin{center}
\includegraphics[width=0.5\textwidth]{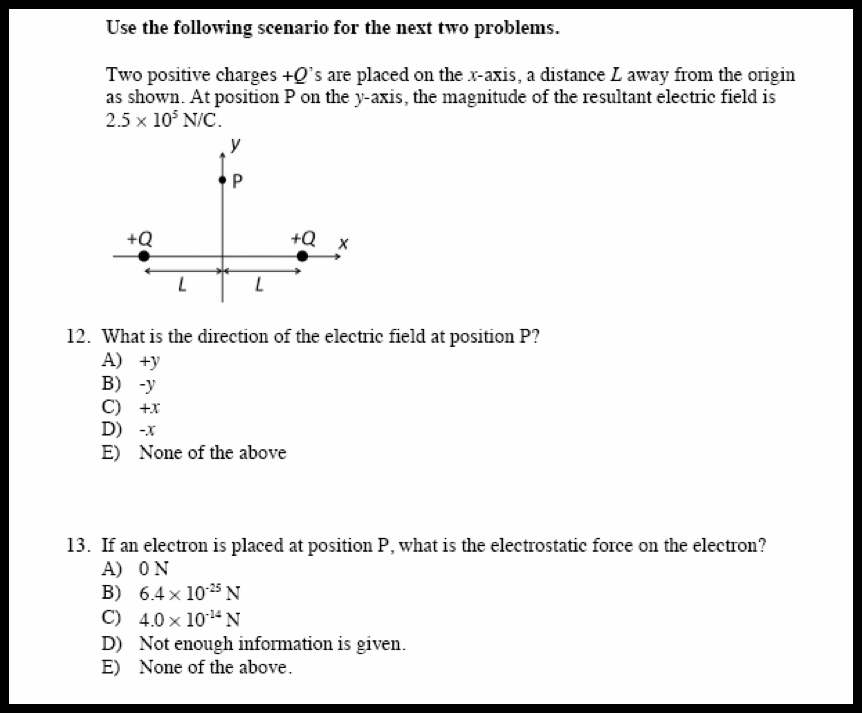}
\caption{The figure shows two exam questions about electric field and force from one of the institutions. Question 12 was categorized as lab related while question 13 was categorized as not lab related.}
\label{fig:Sample1}
\end{center}
\end{figure}

\begin{figure}
\begin{center}
\includegraphics[width=0.5\textwidth]{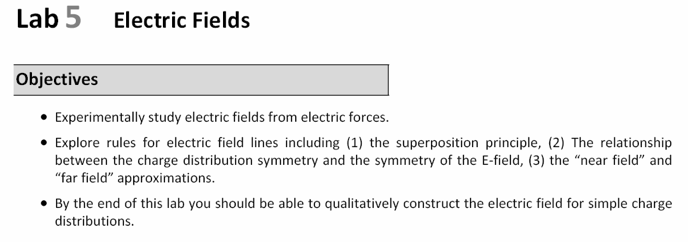}
\caption{The figure shows the list of learning objectives from a lab about electric field and force at the same institution as that used in Fig. \ref{fig:Sample1}.}
\label{fig:Sample2}
\end{center}
\end{figure}

As an example of lab versus non-lab categorizations, Fig. \ref{fig:Sample1} shows a pair of exam questions about electric field and force from one of the institutions and Fig. \ref{fig:Sample2} shows the learning objectives for a lab at that institution about electric field and force. Question 12 in Fig. \ref{fig:Sample1} was categorized as lab related based on the learning objectives. Question 13, however, was categorized as not lab related because the third learning objective in Fig. \ref{fig:Sample2} explains that students will only construct electric fields qualitatively in the lab. The lab does include a brief mention of calculating the electric field when given the resultant force and test charge magnitude. This mention, however, is not phrased as defining the force when given the electric field.  More importantly, the students do not do any example calculations of this sort in the labs.

\begin{figure}
\begin{center}
\includegraphics[width=0.5\textwidth]{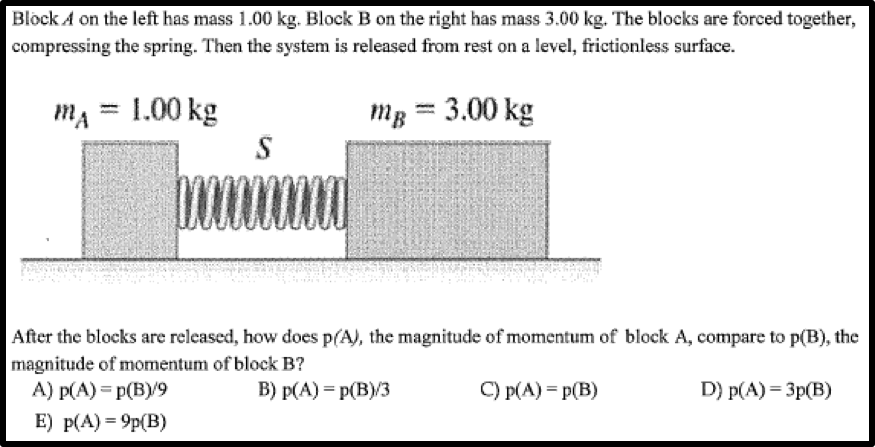}
\caption{The figure shows an exam question about momentum from one of the institutions. It was categorized as conceptual. }
\label{fig:Sample3}
\end{center}
\end{figure}

\begin{figure}
\begin{center}
\includegraphics[width=0.5\textwidth]{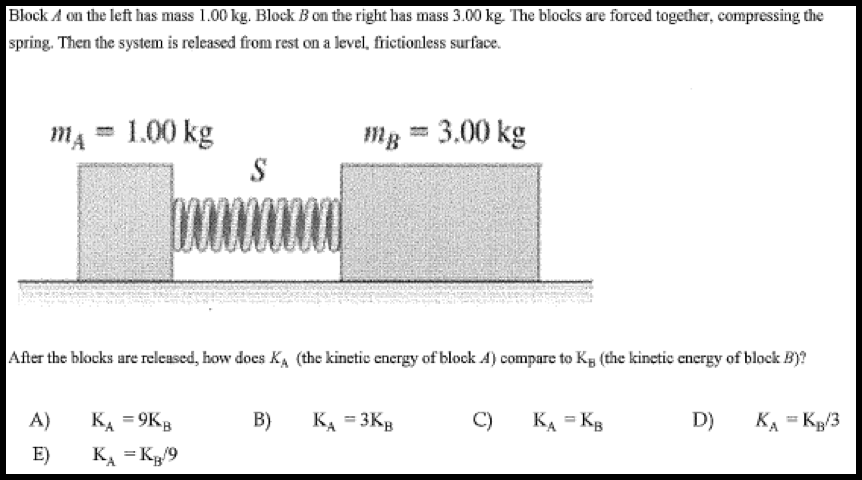}
\caption{The figure shows an exam question about momentum from one of the institutions. It was categorized as calculational.}
\label{fig:Sample4}
\end{center}
\end{figure}

As an example of calculational versus conceptual categorizations, Fig. \ref{fig:Sample3} shows another exam question from an institution, this time about momentum. The question may appear to some to be calculational due to the requirement to determine the ratio of the two momenta. Students need only, however, identify that the system starts from rest (zero initial momentum) before the two blocks separate. Conservation of momentum is sufficient to recognize that the magnitude of the momentum of the two blocks must be the same. In contrast, Fig. \ref{fig:Sample4} shows an exam question from the same institution, also about momentum. This question looks superficially similar to that in Fig. \ref{fig:Sample3}. To answer the question, however, students need both conservation of momentum and conservation of energy. They also need to combine two equations, one with a linear relationship to velocity and the other quadratic. This manipulation of equations is unreasonable to keep track of in a purely `conceptual' way, and so this question was classified as calculational.

\subsection{Statistical analysis}

In the original study \cite{Wieman2015AJP}, the ratio between students' scores on lab-related and non-lab-related items was computed and compared along with the mean lab benefit from the difference scores. We opted to use only the difference scores because the ratio calculation created significant outliers and non-normally distributed data. We used ANOVA methods to compare groups on students' difference scores for the final exams using the $lm$ and $aov$ functions in R \cite{Team2016}. We used repeated-measures ANOVA methods for the midterm exams using the $lmer$ function from the ``Linear Mixed-Effects Models using `Eigen' and S4" package \cite{Bates2015} and $Anova$ function in  using the ``Companion to Applied Regression" package \cite{Fox2011} in R. In all cases, regression analyses were also performed and the same results were obtained. We present only the ANOVA results for simplicity. 

In each ANOVA table below, we provide the degrees of freedom ($df$), the $F$-statistic, the effect size ($\eta_{partial}^2$), and the $p$-value. The effect size measure, $\eta_{partial}^2$, can be interpreted analogously to the coefficient of multiple determination for multiple regression, $R^2$. A $\eta_{partial}^2$ value of 0.3 for Group would mean that 30\% of the between-subjects variance was accounted for by Group. A $\eta_{partial}^2$ around 0.3 is considered relatively large (it can only be between 0 and 1).

For the final exam analyses (both the full exam and only the conceptual questions), we used each students' difference score as the outcome variable, and their enrollment group (lab versus non-lab) and course (to distinguish differences between instructors and exams) as independent variables. In each case, the assumptions of ANOVA were met (continuous dependent variable, categorical independent variables, no significant outliers, normally distributed residuals, independence of cases, and homoscedasticity). The outcome of the analysis for the full final exams and the conceptual questions only are found in Table \ref{tab:FullStats} and Table \ref{tab:FullCStats}, respectively. All three institutions were included in the full-exam analysis, but only two institutions were included in the analysis of the conceptual questions due to availability of data. 

\begin{table}
\caption{The ANOVA analysis of students' difference scores on the final exams at each institution shows no significant effect for enrolling in the lab.}\label{tab:FullStats}
\begin{tabular}{c|cccc}
\hline\hline
Independent variable & $df$ & $F$ &$\eta_{partial}^2$ & $p$\\
\hline
Group & 1 & 2.32 & 0.00005 & .128\\
Instructor & 8 & 38.34 & 0.105 & $<.001^{***}$\\
Residuals & 2615\\
\hline
\multicolumn{5}{r}{$^{***}$ Significant at the $p<.001$ level.}\\
\hline
\hline
\end{tabular}
\end{table}

\begin{table}
\caption{The ANOVA analysis of students' difference scores with conceptual items only on the final exams at each institution shows no significant effect for enrolling in the lab.}\label{tab:FullCStats}
\begin{tabular}{c|cccc}
\hline\hline
Independent variable & $df$ & $F$ &$\eta_{partial}^2$ & $p$\\
\hline
Group & 1 & 1.60 & 0.001 & .206\\
Instructor & 5 & 6.44 & 0.028 & $<.001^{***}$\\
Residuals & 1114\\
\hline
\multicolumn{5}{r}{$^{***}$ Significant at the $p<.001$ level.}\\
\hline
\hline
\end{tabular}
\end{table}

\begin{table}
\caption{The repeated-measures ANOVA analysis of students' difference scores across midterm exams at each institution shows no significant effect for enrolling in the lab.}\label{tab:MidStats}
\begin{tabular}{p{2cm}c|cccc}
\hline\hline
&Independent variable & $df$ & $F$ & $p$\\
\hline
\multirow{5}{2cm}{\rotatebox[origin=c]{90}{\parbox[c]{1cm}{\centering Full midterm exams}}} & Group & 1 & 0.65 & .420\\
&Instructor & 5 & 26.94 & $<.001^{***}$\\
&Exam & 3 & 114.57 & $<.001^{***}$\\
&Exam*Instructor & 8 & 37.56 & $<.001^{***}$\\
&Residuals & 2939\\
\hline
\multirow{5}{2cm}{\rotatebox[origin=c]{90}{\parbox[c]{1.5cm}{\centering Conceptual questions for midterm exams}}} & Group & 1 & 0.10 & .747\\
&Instructor & 4 & 28.67 & $<.001^{***}$\\
&Exam & 2 & 15.95 & $<.001^{***}$\\
&Exam*Instructor & 2 & 7.92 & $<.001^{***}$\\
&Residuals & 2050\\
\hline
\multicolumn{4}{r}{$^{***}$ Significant at the $p<.001$ level.}\\
\hline
\hline
\end{tabular}
\end{table}


For the midterm exams analysis (the full exams and only the conceptual questions), we used repeated-measures ANOVA. This used each students' difference score as the outcome variable, their enrollment group (lab versus non-lab), course (to distinguish differences between instructors and exams), and each exam number as independent variables, and a unique student ID as a random effects variable. Only two of the institutions were included in this analysis due to availability of data. In each case, the assumptions of repeated-measures ANOVA were met (as before, but no longer requiring independence of cases). The outcome of the analysis for the full midterm exams and the conceptual questions only are found in Table \ref{tab:MidStats}. The best-fit models also involved an interaction between the exam and the course and a single random effects intercept for each student (as opposed to an intercept for each student on each exam). The interaction suggests that different courses had variable progressions of student scores on exams, which may be indicative of the variability in instructors, exam difficulty, or student performance over time in different courses.


\end{document}